\documentclass[]{spie}  %>>> use for US letter paper

\renewcommand{\baselinestretch}{1.0} % Change to 1.65 for double spacing
 
\usepackage{amsmath,amsfonts,amssymb}
\usepackage{graphicx}
\usepackage[colorlinks=true, allcolors=blue]{hyperref}
\usepackage{caption}
\usepackage{subcaption}

\title{Steered Mixture-of-Experts Autoencoder Design for Real-Time Image Modelling and Denoising}

\author[a]{Elvira Fleig}
\author[a]{Erik Bochinski}
\author[a]{Thomas Sikora}
\affil[a]{Communication Systems Group, Technische Universität Berlin, Berlin, Germany}

\begin{document} 
\maketitle

\begin{abstract}
Research in the past years introduced Steered Mixture-of-Experts (SMoE) as a framework to form sparse, edge-aware models for 2D- and higher dimensional pixel data, applicable to compression, denoising, and beyond, and capable to compete with state-of-the-art compression methods. To circumvent the computationally demanding, iterative optimization method used in prior works an autoencoder design is introduced that reduces the run-time drastically while simultaneously improving reconstruction quality for block-based SMoE approaches. Coupling a deep encoder network with a shallow, parameter-free SMoE decoder enforces an efficent and explainable latent representation. Our initial work on the autoencoder design presented a simple model, with limited applicability to compression and beyond.\\
In this paper, we build on the foundation of the first autoencoder design and improve the reconstruction quality by expanding it to models of higher complexity and different block sizes. Furthermore, we improve the noise robustness of the autoencoder for SMoE denoising applications. Our results reveal that the newly adapted autoencoders allow ultra-fast estimation of parameters for complex SMoE models with excellent reconstruction quality, both for noise free input and under severe noise. This enables the SMoE image model framework for a wide range of image processing applications, including compression, noise reduction, and super-resolution. 
\end{abstract}


\section{Introduction}
\begin{figure}
\begin{subfigure}[t]{0.13\linewidth}
\includegraphics[width=\linewidth]{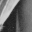}
\caption{Original}
\label{fig:1_orig}
\end{subfigure}
\hfill
\begin{subfigure}[t]{0.13\linewidth}
\includegraphics[width=\linewidth]{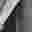}
\caption{\textbf{JPEG}\\ PSNR:$26.33 dB$\\ SSIM: $0.82$}
\label{fig:1_JPEG}
\end{subfigure}%
\hfill
\begin{subfigure}[t]{0.13\linewidth}
\includegraphics[width=\linewidth]{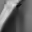}
\caption{\textbf{HEVC}\\PSNR:$26.05 dB$ \\SSIM: $0.77$}
\label{fig:1_HEVC}
\end{subfigure}%
\hfill
\begin{subfigure}[t]{0.13\linewidth}
\includegraphics[width=\linewidth]{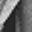}
\caption{\textbf{JPEG2000}\\ PSNR:$29.43 dB$\\ SSIM: $0.87$}
\label{fig:1_JPEG200}
\end{subfigure}%
\hfill
\begin{subfigure}[t]{0.13\linewidth}
\includegraphics[width=\linewidth]{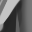}
\caption{\textbf{SMoE}\\ PSNR:$31.66 dB$ \\SSIM: $0.9$}
\label{fig:1_smoe}
\end{subfigure}%
\hfill
\begin{subfigure}[t]{0.13\linewidth}
\includegraphics[width=\linewidth]{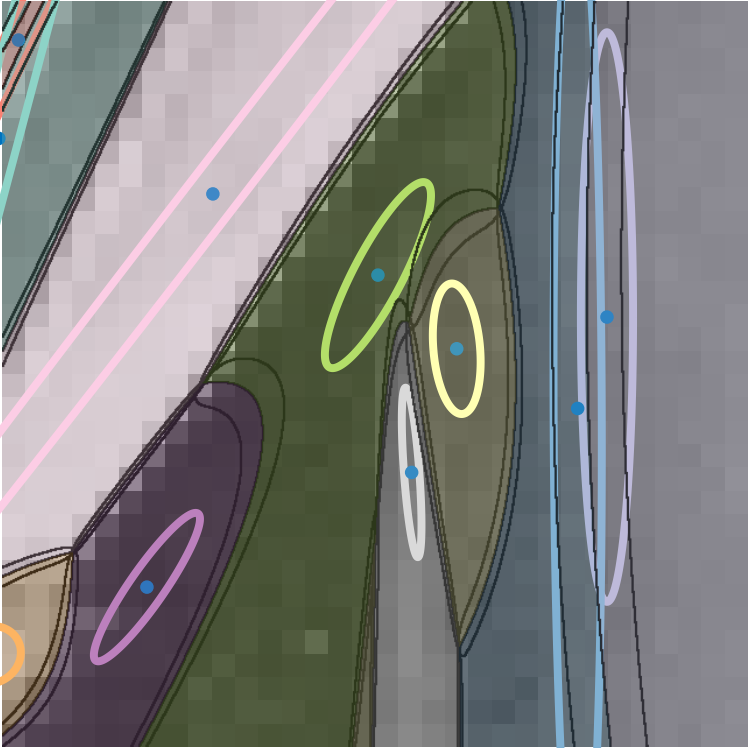}
\caption{SMoE Kernels}
\label{fig:1_kernels}
\end{subfigure}%
\hfill
\begin{subfigure}[t]{0.13\linewidth}
\includegraphics[width=\linewidth]{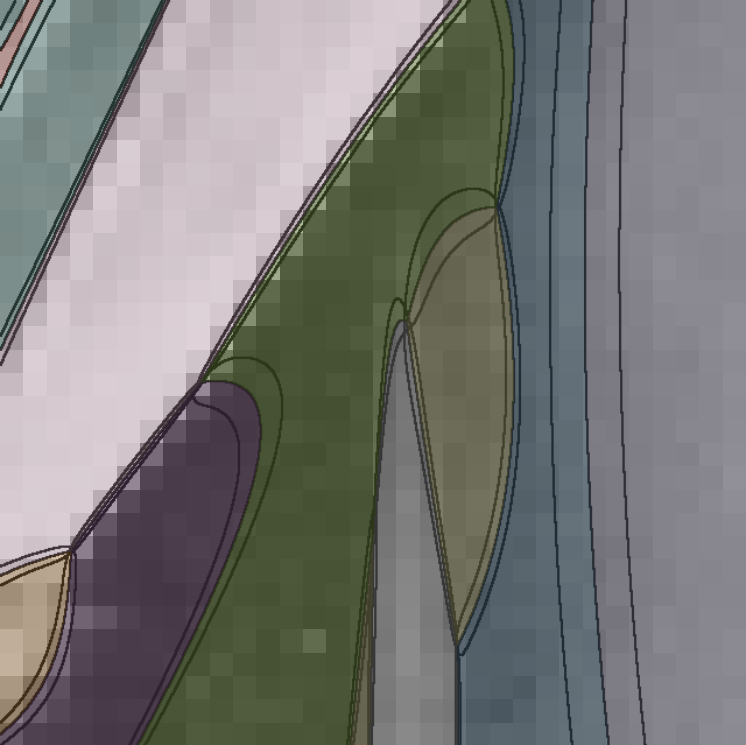}
\caption{SMoE Gates}
\label{fig:1_Gates}
\end{subfigure}%
\vspace{-0.15cm}
\caption{Compression performance at 0.43 bpp\cite{fleig2023edge}}
\label{fig:Comparison_JPEG_HEVC_SMoE}
\vspace{-0.4cm}
\end{figure}

The Steered Mixture-of-Experts (SMoE) framework has been first introduced for coding images \cite{verhack2016universal} with a sparse regression model and has since been further investigated not only for still images but expanded to video \cite{lange2016video} and higher dimensional data such as light field images and video\cite{verhack2019steered}. 
Application fields % nur Applications?
of the framework are in compression, \cite{tok2018mse} \cite{bochinski2018regularized} \cite{jongebloed2018hierarchical} \cite{jongebloed2019quantized} \cite{liu2019image} and recently extended to denoising and super-resolution \cite{AytacSteered}.\\
The SMoE framework describes in-stationarities like edges and smooth transitions in an image through a sparse representation, forming an edge-aware model. Steered kernels model correlations between pixels, following a divide-and-conquer-principle\cite{yuksel2012twenty}. Utilizing its steering properties, excellent edge representations can be obtained. For compression, SMoE parameters are quantized and coded and can be used directly to reconstruct the image, thus maintaining a high interpretability of its parameters. In contrast to other block-based compression approaches like JPEG, SMoE models describe images in the spatial rather than the transformation domain.\\
To optimize the model parameters, Expectation-Maximization (EM) or Gradient-Descent (GD) algorithms can be employed, albeit GD optimization results in higher reconstruction quality \cite{tok2018mse} \cite{bochinski2018regularized}. A further advantage of GD optimization is the ability to apply well-known training techniques explored in neural network based approaches \cite{albawi2017understanding} to the SMoE framework, enabling the use of cost functions like mean-squared-error (MSE) or SSIM. Impressive results rivaling JPEG2000 are presented in the work of Jongebloed et al. \cite{jongebloed2019quantized} and demonstrate the powerful abilities of the regression model.\\
The basis for the SMoE image model is an edge-aware, continuous non-linear regression function, capable of modeling smooth and sharp transitions in an image without blocking or ringing artifacts, which can be seen in JPEG-like encoders. Fig. \ref{fig:Comparison_JPEG_HEVC_SMoE} depicts the functionality of the model and illustrates the reconstruction results in contrast to other image compression standards. The SMoE reconstruction displays outstanding edge quality, simultaneously preserving smooth transitions in the image. At the same bit rate, both PSNR and SSIM are improved compared to JPEG, JPEG2000, and HEVC-Intra. By creating an edge-aware model of the image, light noise in the original image is also removed, showing its native denoising capabilities.\\
The foundation of our model are gaussian kernels, displayed in Fig. \ref{fig:1_kernels}. The locations and steering properties harvest the correlation between pixels and form the gating function, depicted in Fig. \ref{fig:1_Gates}, which support the edge-aware properties of the model. Sharp edges are modeled with minimal overlapping gates, and smooth transitions by vastly overlapping ones. SMoE thus forms a description of edges and smooth transitions emphasizing important features in an image.\\
Describing images through a SMoE model allows for modeling beyond the regular N-dimensional pixel grid and can be applied to arbitrarily shaped data like point clouds or irregularly sampled imagery. Additionally it can be extended to N-dimensional signals including light-fields \cite{verhack2017steered} and video data \cite{lange2016video}. Once built, the continuous SMoE model allows resampling imagery to any resolution in time and space, which includes edge-aware super-resolution and motion-interpolation.\\
A major drawback of the SMoE model is the iterative nature of GD optimization, resulting in high processing times to obtain well-optimized SMoE parameters which make this approach unfeasible for real-time applications. To circumvent the interative training method, an autoencoder design was introduced in our previous work\cite{fleig2023edge} enabling the estimation of SMoE parameters with run-time savings by a factor of $500-1,000$ compared to the work of Tok et al.\cite{tok2018mse}, while achieving comparable reconstruction qualities. Combining a deep encoder network with a shallow SMoE decoder enforces the encoder to provide SMoE parameters in the latent space, ready for compression and coding, simultaneously maintaining full compatibility with the established SMoE framework. However, the initial autoencoder design\cite{fleig2023edge} only supports radial kernels without steering capabilities, which provides narrow applicability for further SMoE applications.\\
In this work, we present an improved framework of the autoencoder design extended to model complex image patterns and adapted to serve promising SMoE denoising applications. First, steering properties are enabled resulting in the ability to model image structures of higher complexity and consequently improve reconstruction quality. Additionally, the effect of smaller block sizes on both the reconstruction quality in terms of PSNR and SSIM and run time is examined, leading to a higher kernel density to further improve the reconstruction quality. In previous work SMoE models showed promising results for image denoising applications \cite{AytacSteered}, unfortunately with high computational demands. In this paper, we investigate SMoE-AE models trained on noisy data for applications in SMoE denoising algorithms.
\section{Edge-aware Steered Mixture-of-Experts Model}
The scope of Steered Mixture-of-Experts regression is to optimize the underlying regression function Eq. \ref{eq:Grundgleichung} to determine the luminance value $y_p(\underline{x})$ of each position $\underline{x}$ in an image. The main components of this function are the experts $m_i(\underline{x})$ and gating function $w_i(\underline{x})$, which form a weighted sum over all $K$ kernels:
\begin{align}
    y_p(\underline{x})=\sum_{i=1}^{K} m_i(\underline{x}) \cdot w_i(\underline{x})
    \label{eq:Grundgleichung}
\end{align}
While this work focuses on constant experts $m_i(\underline{x})=m_i$, hyperplanes \cite{verhack2016universal} or polynomials are equally possible. In Eq. \ref{eq:Grundgleichung} the experts are weighted by the gating functions $w_i(\underline{x})$. \\
In our work, each gating function $w_i(\underline{x})$ is defined as a soft-max function consisting of $K$ gaussian kernels $\mathcal{K}$:
\begin{align}
    w_i(\underline{x})=\frac{\mathcal{K}(\underline{x},\underline{\mu}_i,\underline{\Sigma}_i)}{\sum_{j=1}^{K}\mathcal{K}(\underline{x},\underline{\mu}_j,\underline{\Sigma}_j)}
\end{align}
More specifically:
\begin{align}
   \mathcal{K}(\underline{x},\underline{\mu}_i,\underline{A}_i)=exp[-\frac{1}{2}(\underline{x}-\underline{\mu}_i)^T \underline{A}_i\underline{A}_i^{T} (\underline{x}-\underline{\mu}_i)]
   \label{eq:Kernelgleichung}
\end{align}
with $\underline{\mu}_i$ being the center position of each kernel and $\underline{A}_i$ the steering properties. The covariance matrix of the gaussian kernel $\underline{\Sigma}_i$ is thus given as the inverse Cholesky decomposition, to stabilize training by omitting matrix inversion and ensure positive semidefiniteness. $\underline{A}_i$ is given as a lower triangular matrix with the following dimensions:
\begin{align}
   \underline{A}_{i} =
   \begin{pmatrix}
 a_{11} & 0\\
  a_{21} &  a_{22}
\end{pmatrix}
\label{eq:Steeringparams}
\end{align}
However, the dimension of $\underline{A}_i$ can be extended accordingly for higher dimensional data.\\
Together with the experts $m_i$, the center positions $\underline{\mu}_i$ in each dimension of the image and the steering parameters in $\underline{A}_{i}$ need to be estimated from the pixel data, resulting in 6 parameters per kernel.\\
A simpler representation was used in prior works \cite{tok2018mse} \cite{fleig2023edge} using radial kernels instead of steered kernels. This drastically reduces the number of required parameters, while still maintaining a good representation of edges and smooth transitions in images \cite{tok2018mse}. When assigning the same bandwidth for all kernels $B$, the regression function Eq. \ref{eq:Grundgleichung} simplifies to:
\begin{align}
   y(\underline{x})=\sum_{i=1}^{K} m_i\cdot\frac{exp(-\mathcal{B}||\underline{x}-\underline{\mu}_i||^2)}{\sum_{j=1}^{K}exp(-\mathcal{B}||\underline{x}-\underline{\mu}_j||^2)}
   \label{eq:OPGleichung}
\end{align}
Employing this reconstruction equation the needed SMoE parameters reduce to 3 per kernel. The SMoE model parameters can be effectively optimized using the gradient descent (GD) method as proposed in \cite{tok2018mse} \cite{bochinski2018regularized}, which outperforms the EM Algorithm in reconstruction quality\cite{verhack2016universal}. Optimization using GD defines a loss function $\mathcal{L}$ to calculate e.g. the mean squared error over all pixels $N$.
\begin{align}
   \mathcal{L}:=\frac{1}{N}\sum_{n=1}^N(y_n(\underline{x})-\sum_{i=1}^K m_i(\underline{x}_n)\cdot w_i(\underline{x}_n))^2
\end{align}
%\vspace{-0.3cm}
\begin{align}
    \underset{m_i,\underline{\mu}_i,\underline{A}_{i}}{\mathrm{arg\,min}} \{\mathcal{L}\}
\end{align}
 The drawback of models trained using GD, to which we refer as SMoE-GD, is the high dependency on the initialization of the center position $\underline{\mu}_i$ which can lead to subpar results by finding only suboptimal minima. Furthermore, to achieve satisfactory results many training iterations are necessary causing a high computational cost. This makes SMoE-GD unfeasible for many real-time applications.\\
%Our previously introduced autoencoder design (SMoE-AE) restricted to radial kernels with the capability to achieve comparable or better results in a fraction of the time compared to SMoE-GD \cite{fleig2023edge}, as shown further below. The concept is depicted in Fig. \ref{fig:structure_AE}.
%It is comprised of a deep encoder network in conjunction with a shallow SMoE decoder. The bottleneck layer at the output of the encoder network yields the SMoE parameters and can be used directly in Eq. \ref{eq:OPGleichung} to reconstruct the image. 

\section{Autoencoder Design for Steered Mixture-Of-Experts}
Previous work \cite{fleig2023edge} introduced the autoencoder design for radial kernels and showed its efficiency in comparison to SMoE-GD. In this work, we further develop SMoE-AE, to be applicable to the aforementioned SMoE applications, to reduce the run time, and make the SMoE model feasible for real-time applications. 
%To increase the reconstruction quality further and adjust the autoencoder to achieve better results in the respective application fields, additional design changes have been made and will be presented in this work. 
\begin{figure}
\centering
\includegraphics[width=0.7\linewidth]{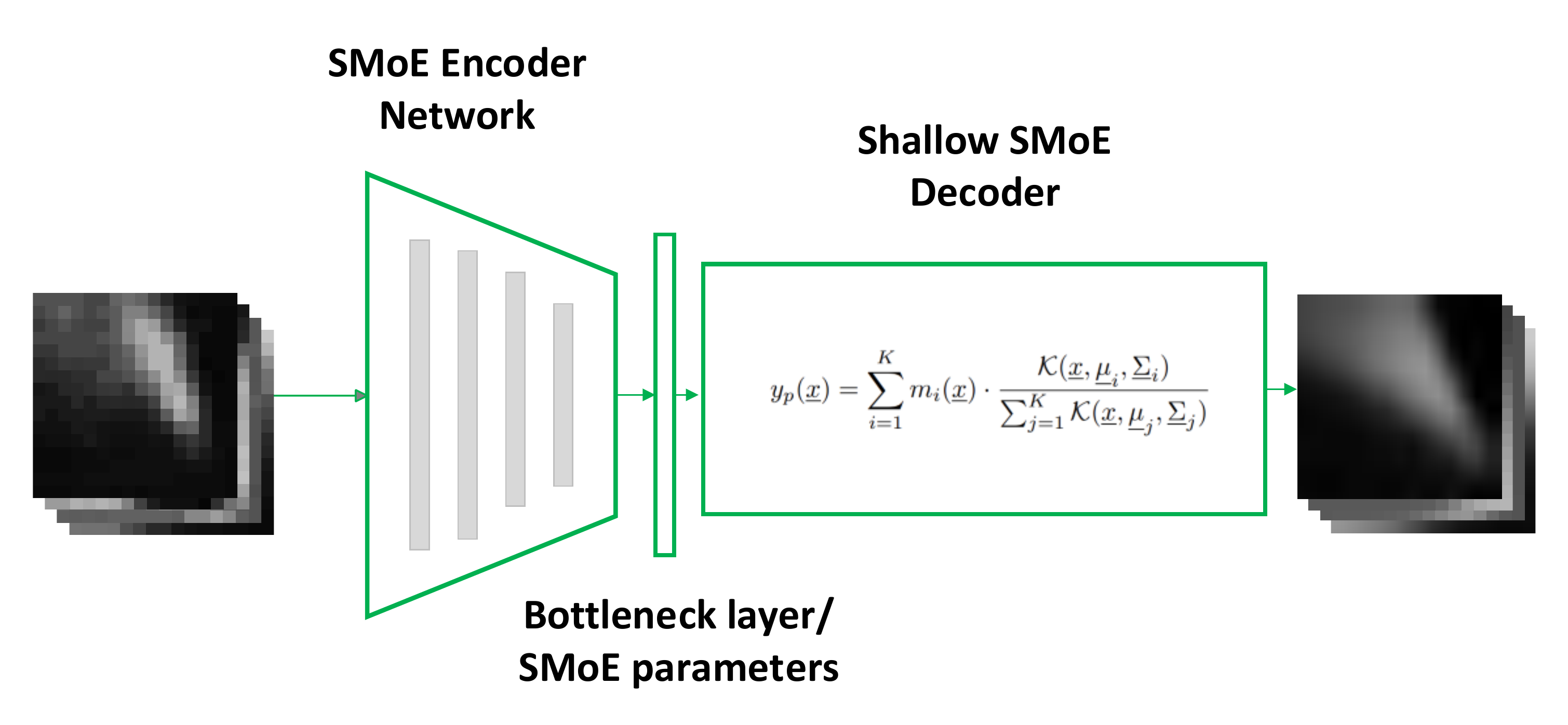}
\caption{SMoE Autoencoder}
\label{fig:structure_AE}
%\vspace{-0.3cm}
\end{figure}
The basic concept of SMoE-AE was presented in Fleig et al.\cite{fleig2023edge}. Fig. \ref{fig:structure_AE} illustrates the approach and depicts the principal structure of the autoencoder design. It deviates from the conventional autoencoder design \cite{cheng2018deep}\cite{meng2017research}, by combining a deep encoder network with a shallow untrainable SMoE decoder, which performs the SMoE reconstruction based on Eq. \ref{eq:Grundgleichung}. 
The deep encoder network is trained on ground truth images in an end-to-end process, coupled with the SMoE decoder. By combining the SMoE decoder in an autoencoder architecture, we ensure the encoder predicts feasible SMoE parameters in the latent space and simultaneously enables loss functions like MSE or SSIM to optimize rate-distortion. 
The first SMoE-AE was designed to be comparable directly with the work of Tok et al.\cite{tok2018mse} and is the basis for this work. In \cite{fleig2023edge} the bottleneck layer was restricted to yield parameters for $K=4$ radial kernels, while the Bandwidth $B$ was provided as a hyperparameter.
The input data was comprised of $16\times 16$ pixel greyscale blocks and was designed to reconstruct blocks of the same dimensions. The reader is referred to \cite{fleig2023edge} to see the architectural structure of the basic encoder network, in addition to training specifications. 
This work aims to build upon the foundation of the basic SMoE-AE design and expands the framework to reconstruct higher complexity image data and improve noise robustness.\\
The great advantage of SMoE is the aforementioned ability to form an edge-aware model in the spatial domain, which is best utilized by employing steering properties to the kernels. 
%Therefore these kernels can steer along edges and model them in impressive detail, without causing blocking or ringing artifacts like other well-known compression methods. 
In this work we enable steering properties via the Cholesky decomposition in Eq. \ref{eq:Kernelgleichung} in the SMoE decoder, which additionally ensures $\underline{A}_i$ to be a lower triangular matrix. By expanding the number of parameters needed to satisfy Eq. \ref{eq:Kernelgleichung} with three additional parameters/kernel, the latent space output of the encoder network is increased accordingly. We restrict the number of Kernels to $K=4$, thus resulting in a 24-dimensional output. To account for the increased complexity of the latent space output, additional layers in the encoder network are added. The full structure is depicted in Tab. \ref{tab:Architecture}.\\
\begin{table}[]
\begin{center}
\caption{Architectural structure of SMoE Encoder}
\label{tab:Architecture}
\begin{tabular}[t]{l l l l}

\hline
\multicolumn{2}{c}{Convolutional Layer}& \multicolumn{2}{c}{Dense Layer}\\
\hline
Quantity & 7 & Quantity 6\\
Filter&16,32,64,128,256,512,1024&Output Dimension & 1024,512,256,128,64,24\\
Padding&Same&&\\
Kernel Size&3x3&&\\
\hline

\end{tabular}
\end{center}
\end{table}
To improve the reconstruction quality and explore the applicability of SMoE-AE to various use cases of SMoE we train our encoder network on a smaller block size of $8\times8$ pixel/block. The general framework of the encoder remains equal, maintaining the complexity demands on the network, and simultaneously quadrupling the kernel density over the whole image. Smaller block sizes of dimension $8\times 8$ are beneficial to perform denoising on low to medium levels of noise \cite{dabov2007image} and are i.e. used in BM-SMoE and S-SMoE algorithm\cite{AytacSteered} for denoising. 
\begin{figure}
\centering
\includegraphics[width=0.9\linewidth]{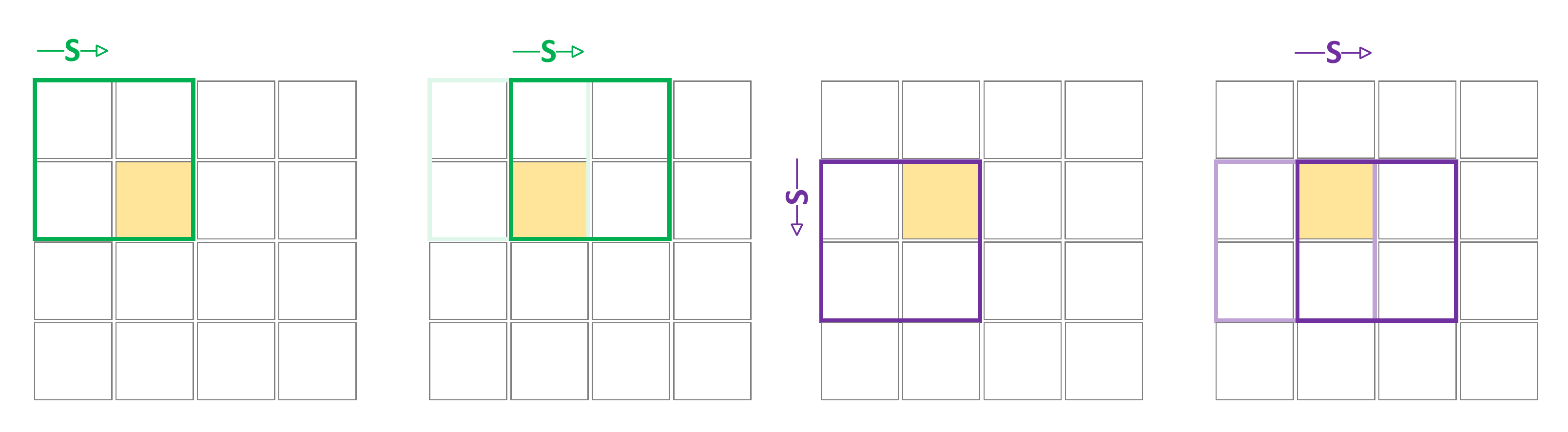}
\caption{Principal functionality of S-SMoE, for a window of $2x2$ resulting in four hypotheses for the yellow pixel point.}
\label{fig:S-Smoe_basic}
%\vspace{-0.3cm}
\end{figure}
Additionally BM-SMoE and S-SMoE have great potential in super-resolution applications \cite{AytacSteered}. Its basic concept is depicted in Fig. \ref{fig:S-Smoe_basic}. To employ S-SMoE a sliding window with dimensions $N\times N$ pixels and a step size $S$ pixels is moved across the image. For each extracted block SMoE parameters are predicted. Using a step size $S$ smaller than the dimensions of the sliding window $N$ results in overlapping blocks, thus receiving multiple hypotheses of reconstructions at every single pixel point of the image, depicted in yellow in Fig. \ref{fig:S-Smoe_basic}. SMoE regression is performed for every extracted block and repeating this process for every overlapping block results in numerous hypotheses for each pixel of the image, each deviating slightly due to the sparse representation of the model. 
Averaging all possible blocks results in a reliable representation for each pixel of the image.  With a growing number of hypotheses, a higher sampling of all edges in an image is performed.
Unfortunately, optimizing SMoE parameters through SMoE-GD for a small step size is computationally unfeasible, posing an optimal use-case for SMoE-AE run-time acceleration.\\ 
\begin{table}[]

\begin{center}
\caption{Comparison of SMoE-GD and SMoE-AE for radial and steered Kernels and 16x16 blocks}
\label{tab:Steering}
\begin{tabular}[t]{l l l l l l l l l}

\hline
& \multicolumn{4}{c}{SMoE-GD}& \multicolumn{4}{c}{SMoE-AE}\\
&\multicolumn{2}{c}{Radial Kernels}&\multicolumn{2}{c}{Steered Kernels}&\multicolumn{2}{c}{Radial Kernels}&\multicolumn{2}{c}{Steered Kernels}\\
\hline
Sequence&PSNR [dB]&SSIM &PSNR [dB]&SSIM&PSNR [dB]&SSIM&PSNR [dB]&SSIM\\
%&[dB&&[bpp]&dB&&[bpp]&dB&\\
Baboon&22.91&0.42&23.63&0.63&22.95&0.56&23.59&0.63\\
Boats&25.46&0.69&26.28&0.72&25.65&0.70&26.64&0.74\\
Bridge&22.80&0.54&23.30&0.59&22.91&0.55&23.56&0.60\\
Cameraman&27.08&0.85&28.48&0.86&27.79&0.85&29.17&0.88\\
Elaine&30.63&0.75&30.86&0.86&30.34&0.73&31.02&0.75\\
Lena&28.83&0.82&29.80&0.84&29.05&0.82&29.91&0.84\\
Livingroom&25.19&0.67&25.97&0.71&25.55&0.69&26.46&0.73\\
Peppers&29.50&0.79&29.94&0.79&29.56&0.78&30.52&0.80\\
Kodak&25.77&0.70&26.08&0.71&25.94&0.70&26.92&0.74\\
\hline
Encoding Time [s]&\multicolumn{2}{c}{296.29}&\multicolumn{2}{c}{347.36}&\multicolumn{2}{c}{0.25}&\multicolumn{2}{c}{0.44}\\
Decoding Time [s]&\multicolumn{2}{c}{0.02}&\multicolumn{2}{c}{0.22}&\multicolumn{2}{c}{0.02}&\multicolumn{2}{c}{0.22}\\
Encoding RTG&\multicolumn{2}{c}{1185}&\multicolumn{2}{c}{789}&&&&\\
%&0.16&25.15&0.64&-&-&-&0.16&\textbf{25.54}&\textbf{0.68}\\
\hline

\end{tabular}
\end{center}
\end{table}
\section{Experiments}
The final encoder network comprises over $74M$ trainable parameters in total for $8\times 8$ blocks and over $275M$ for $16\times 16$ input blocks. The increased amount of trainable parameters for bigger block sizes result in higher training times which consumed around $20 h$ for $8\times 8$ blocks and $60 h$ for $16\times 16$ blocks on an \textit{Nvidia GTX 1070 Ti}. Regardless of block size, the innermost bottleneck layer of the encoder network yields the expert values $m_i$, the kernel center values $\underline{\mu}_i$ and the steering parameters $a_1$ to $a_3$ for every kernel and can be used directly to solve the regression function in Eq. \ref{eq:Grundgleichung}. 
To train the encoder network the mobile subset of the clic dataset \cite{CLIC2020} was used. The subset contains $1048$ training images and $61$ validation images, the testing was made on well-known $512\times512$ test images\cite{USCImageDatabase} and  all $24$ images of the \textit{Kodak PhotoCD} dataset \cite{KodakPhotoCd}. Therefore all images were converted to grey levels and the pixel values were normalized to the $[0...1]$ domain. Then the images were cropped to $1024\times 1024$ and divided into $16\times 16$ pixel or $8\times 8$ pixel blocks respectively, resulting in a total of over $4$ million for $16\times 16$ pixel blocks and over $17$ million for $8\times 8$ pixel blocks.\\
It is necessary to train the network for every block size separately since SMoE-AE adapts to model more detailed structures in smaller blocks, and in larger blocks to more coarse structures. This results in a loss of quality when predicting SMoE parameters for $8\times8$ pixel blocks with a network trained on $16\times16$ pixel blocks.\\
Test images were preprocessed equally, except the dimensions of the images remained unchanged. For training with noise, noise was added to each block, regardless of train or test data.
The original block served as the training target. During training, the blocks were shuffled after every training epoch to encourage the best possible generalization of the encoder network.\\
Regarding the innermost layer, the parameters were split into expert values $m_i$ and kernel center values $\underline{\mu}_i$, which were restricted to $[0...1]$, while the steering parameters $a_1$ to $a_3$ remained unrestricted to achieve full steering capabilities. These parameters ranged for most images between around $-50$ and $+50$.\\
The training was performed over $30$ epochs using Adam Optimizer with default parameters, a batch size of $64$, and a constant learning rate of $l=5\cdot10^{-5}$. The loss was calculated as the MSE of the original block to the reconstruction. Rectified linear unit (ReLU) was used as the activation function, except in the innermost layer a linear activation function was used to enable an unrestricted range of the steering parameters.\\ 
After training the encoder and decoder were uncoupled and the pre-trained encoder network was used to predict SMoE parameters for any image block.\\
The primary goal of SMoE-AE is to reduce the run-time in estimating SMoE parameters while maintaining the reconstruction quality in terms of PSNR and SSIM of SMoE-GD. Consequently, we used SMoE-GD as a benchmark to compare the capabilities of SMoE-AE. To objectively compare run-time gains (RTG) of SMoE-AE to SMoE-GD, we ensure convergence of SMoE-GD by performing its optimization for $5,000$ iterations.\\
\begin{figure}

\begin{subfigure}[t]{0.24\linewidth}
\includegraphics[width=\linewidth]{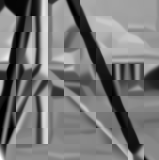}
\caption{Radial kernel\\$16\times16$ blocks}
\label{fig:Cameraman_radial}
\end{subfigure}
\centering
\begin{subfigure}[t]{0.24\linewidth}
\includegraphics[width=\linewidth]{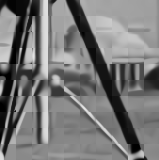}
\caption{Steered kernel\\$16\times16$ blocks}
\label{fig:Cameraman_steered}
\end{subfigure}%
\centering
\begin{subfigure}[t]{0.24\linewidth}
\includegraphics[width=\linewidth]{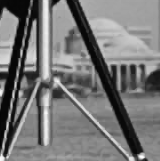}
\caption{Steered kernel\\$8\times8$ blocks}
\label{fig:Cameraman_steered_8x8}
\end{subfigure}%
\centering
\begin{subfigure}[t]{0.24\linewidth}
\includegraphics[width=\linewidth]{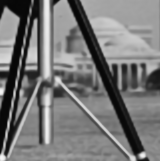}
\caption{S-SMoE with $S=1$\\$8\times8$ blocks}
\label{fig:Cameraman_steered_shift1}
\end{subfigure}%
\caption{Crop of test image \textit{Cameraman} reconstructed with radial and steered kernels and different block sizes}
\label{fig:Kernel_crops}

\end{figure}
\subsection{Supporting variable block-sizes and model complexities}
Tab. \ref{tab:Steering} provides results comparing SMoE-GD and SMoE-AE on $16\times 16$ blocks. For radial kernels both SMoE-GD and SMoE-AE are comparable in results regarding PSNR and SSIM, achieving only marginal differences in reconstruction quality.
SMoE-GD achieves better results for the image \textit{Elaine}, with $0.29$dB higher PSNR and $0.2$ higher SSIM, while the test image \textit{Cameraman} converges to a PSNR $0.71$dB lower than SMoE-AE. This observation affirms the assumption, that SMoE-GD is highly dependable on the initialization of the parameters, which can lead to difficulties in finding global optima but can occasionally form excellent results. Even though the results of SMoE-AE and SMoE-GD are very similar, the encoding time differs drastically. 
On average approx. $300s$ per image are necessary for the SMoE-GD encoder to converge and to achieve comparable results to SMoE-AE, which predicts SMoE parameters in $0.25s$/image, resulting in an encoding RTG by a factor of $>1,000$.
This tremendous gain in run time is remarkable, given the preserved reconstruction quality. By maintaining the shallow properties at the decoder network the reconstruction of the image from SMoE Parameters is as equally fast as SMoE-GD.\\
Tab. \ref{tab:Steering} depicts additionally the reconstruction results for $K=4$ steered kernels. Adding steering capabilities leads to the doubling of parameters in the latent space. However, the autoencoder framework can easily adapt to these increased demands and is trained to predict the additional parameters with a high reconstruction quality. Regardless of the optimization procedure, the PSNR improved substantially, grounded in the improved ability of kernels to steer along edges and model more complex patterns in a block. This is depicted in Fig.\ref{fig:Cameraman_radial} and Fig. \ref{fig:Cameraman_steered} for a crop of the test image \textit{Cameraman}. The struts of the tripod are more detailed and appear sharper using steered kernels compared to radial kernels. Such fine details can be better represented with steered kernels, as they can steer in a narrow line, thus creating a sharp line in the image, with sharp transitions on both sides. Compared to SMoE-GD the adapted SMoE-AE autoencoder achieves $0.8$dB reconstruction gain on the \textit{Kodak} data set on average while speeding up the process by a RTG factor of $750$. The average time for reconstructing a $512\times 512$ image reduced from $>5$ minutes remarkably to $< 1$ second. \\ 
\begin{table}[]
\centering
\caption{Reconstruction Quality comparison on 16x16 pixel blocks and 8x8 pixel blocks}
\label{tab:smaller_blocks}
\begin{tabular}[t]{l l l l l l l l l}

\hline
& \multicolumn{4}{c}{SMoE-GD}& \multicolumn{4}{c}{SMoE-AE}\\
&\multicolumn{2}{c}{16x16}&\multicolumn{2}{c}{8x8}&\multicolumn{2}{c}{16x16}&\multicolumn{2}{c}{8x8}\\
\hline
Sequence&PSNR [dB]&SSIM &PSNR [dB]&SSIM&PSNR [dB]&SSIM&PSNR [dB]&SSIM\\
%&[dB&&[bpp]&dB&&[bpp]&dB&\\
Baboon&23.63&0.63&28.50&0.89&23.69&0.63&29.26&0.91\\
Boats&26.28&0.72&30.84&0.87&26.64&0.74&32.04&0.89\\
Bridge&23.30&0.59&26.49&0.81&23.56&0.60&27.52&0.86\\
Cameraman&28.48&0.86&35.39&0.96&29.17&0.88&35.43&0.97\\
Elaine&30.86&0.86&32.79&0.98&31.02&0.75&33.25&0.83\\
Lena&29.80&0.84&34.40&0.92&29.91&0.84&35.37&0.93\\
Livingroom&25.97&0.71&30.18&0.87&26.46&0.73&32.06&0.91\\
Peppers&29.94&0.79&33.24&0.86&30.52&0.80&34.07&0.88\\
Kodak&26.08&0.71&29.35&0.85&26.92&0.74&31.68&0.91\\
\hline
Encoding Time [s]&\multicolumn{2}{c}{347.36}&\multicolumn{2}{c}{346.56}&\multicolumn{2}{c}{0.44}&\multicolumn{2}{c}{0.7}\\
Decoding Time [s]&\multicolumn{2}{c}{0.22}&\multicolumn{2}{c}{0.24}&\multicolumn{2}{c}{0.22}&\multicolumn{2}{c}{0.24}\\
Encoding RTG&\multicolumn{2}{c}{789}&\multicolumn{2}{c}{495}&&&\\
\hline
\end{tabular}

\end{table}
Reducing the block size to $8\times 8$ pixel blocks leads to a four times higher density of kernels per block while maintaining the general framework of the encoder network and parameters per block.
The results depicted in Tab. \ref{tab:smaller_blocks} compare the reconstruction quality of $16\times 16$ and $8\times 8$ pixel blocks for SMoE-GD and SMoE-AE with steered kernels.
It is expected to achieve higher reconstruction quality with smaller blocks, as is true for both optimization approaches. The quality gain for SMoE-AE is higher compared to SMoE-GD since the network was able to optimize the kernel positions to the smaller kernel size and higher density. 
%The increased number of parameters in need of optimization results in a slower conversion time for SMoE-GD, thus leading to lower PSNR and SSIM values for all test images.
The increased number of parameters in total required to model all blocks in the image results in a slower convergence for SMoE-GD, thus leading to a lower gain in reconstruction quality for all test images. On the \textit{Kodak} data set with $8\times 8$ blocks the SMoE-AE improves the reconstruction quality by a remarkable $2$dB with again drastically reduced run time. For SMoE-GD to achieve comparable results to SMoE-AE around $7,500$ iterations are necessary, leading to an encoding RTG of $>1,000$. A visual comparison of $16\times16$ to $8\times8$ pixel blocks is depicted in Fig. \ref{fig:Cameraman_steered} and Fig. \ref{fig:Cameraman_steered_8x8}. Blocking artifacts appear reduced and edges stretching over multiple blocks are continuous. Small features are displayed in high detail due to the high density of the kernels, leading to an overall higher-quality reconstruction.\\
\begin{table}[]
\centering
\caption{S-SMoE with shift of $S=1$,$S=4$ and $S=8$}
\label{tab:s-smoe}
\begin{tabular}[t]{l l l l l l l}

\hline
& \multicolumn{6}{c}{S-SMoE with SMoE-AE}\\
&\multicolumn{2}{c}{$S=1$}&\multicolumn{2}{c}{$S=4$}&\multicolumn{2}{c}{$S=8$}\\
\hline
Sequence&PSNR [dB]&SSIM &PSNR [dB]&SSIM&PSNR [dB]&SSIM\\
%&[dB&&[bpp]&dB&&[bpp]&dB&\\
Baboon&32.54&0.96&31.35&0.94&29.26&0.91\\
Boats&34.43&0.92&33.69&0.91&32.04&0.89\\
Bridge&29.37&0.9&28.79&0.89&27.52&0.86\\
Cameraman&40.23&0.98&38.80&0.98&36.43&0.97\\
Elaine&34.11&0.85&33.86&0.85&33.25&0.83\\
Lena&37.76&0.95&37.01&0.95&35.37&0.93\\
Livingroom&34.56&0.94&33.73&0.93&32.06&0.91\\
Peppers&35.18&0.89&34.83&0.89&34.07&0.88\\
Kodak&34.03&0.94&33.30&0.93&31.68&0.91\\
\hline
Encoding Time [s]&\multicolumn{2}{c}{41.60}&\multicolumn{2}{c}{3.00}&\multicolumn{2}{c}{0.70}\\
Decoding Time [s]&\multicolumn{2}{c}{13.95}&\multicolumn{2}{c}{0.88}&\multicolumn{2}{c}{0.24}\\
%&0.16&25.15&0.64&-&-&-&0.16&\textbf{25.54}&\textbf{0.68}\\
\hline
\end{tabular}

\end{table}
Using SMoE-GD in S-SMoE requires computational run-times of 6-7 hours on one greyscale $512\times 512$ pixel image with $S=1$, while parameter prediction with SMoE-AE allows for full image results in a viable time as depicted in Tab. \ref{tab:s-smoe}. Using a shift of $S=1$ results in $255k$ blocks for a $512\times 512$ pixel image, a computationally intensive task even combined with the fast prediction times of SMoE-AE. 
Incorporating SMoE-AE results in an encoding RTG of around $500$ in comparison to SMoE-GD. Apart from run-time, the advantages of S-SMoE are reported in Tab. \ref{tab:s-smoe}. $S=8$ equates to non-overlapping blocks and is equal to native SMoE-AE. A step size of $S=4$ results in an overlap of a half block with each step, therefore four hypotheses can be made for every pixel, except for pixels at the border of the image, since no additional padding is implemented. Building the average over four hypotheses already leads to great improvements of $1.5$dB on average in PSNR and $0.2$ in SSIM. Additionally, averaging over a step size smaller than the block size leads to the reduction of blocking artifacts, that are visible otherwise. This effect is displayed in Fig. \ref{fig:Cameraman_steered_shift1} for $S=1$. With a step size of $S=1$ a total of $64$ hypotheses are established for each pixel point in an image, thus improving the result further by $0.7$dB in PSNR and $0.1$ in SSIM compared to $S=4$, without blocking artifacts. These capabilities can be utilized in applications like denoising.\\
\subsection{Noise Robustness}
\begin{table}[]
\centering
\caption{Comparison of SMoE-GD, SMoE-AE, and SMoE-AE$_{noisy}$ with added speckle noise of variance $\delta^2=0.01$}
\label{tab:added_noise}
\begin{tabular}[t]{l l l l l l l l l}

\hline
&\multicolumn{2}{c}{Noisy Original}& \multicolumn{2}{c}{SMoE-GD}& \multicolumn{2}{c}{SMoE-AE}& \multicolumn{2}{c}{SMoE-AE$_{noisy}$}\\
\hline
Sequence&PSNR [dB]&SSIM&PSNR [dB]&SSIM &PSNR [dB]&SSIM&PSNR [dB]&SSIM\\
%&[dB&&[bpp]&dB&&[bpp]&dB&\\
Baboon&25.53&0.73&27.16&0.82&27.13&0.81&27.10&0.82\\
Boats&25.38&0.62&28.81&0.77&28.67&0.75&29.60&0.81\\
Bridge&26.35&0.81&25.80&0.77&26.24&0.79&26.07&0.78\\
Cameraman&25.64&0.56&31.26&0.78&30.32&0.74&32.60&0.90\\
Elaine&25.09&0.56&29.94&0.78&28.98&0.66&30.97&0.75\\
Lena&25.69&0.56&31.02&0.78&30.09&0.73&32.16&0.81\\
Livingroom&25.95&0.67&28.61&0.79&28.99&0.79&29.51&0.83\\
Peppers&25.97&0.61&30.56&0.76&29.86&0.73&31.49&0.81\\
Kodak&26.98&0.67&28.10&0.77&28.81&0.77&29.03&0.80\\
\hline

\end{tabular}

\end{table}
\begin{figure}
\begin{subfigure}[t]{0.24\linewidth}
\includegraphics[width=\linewidth]{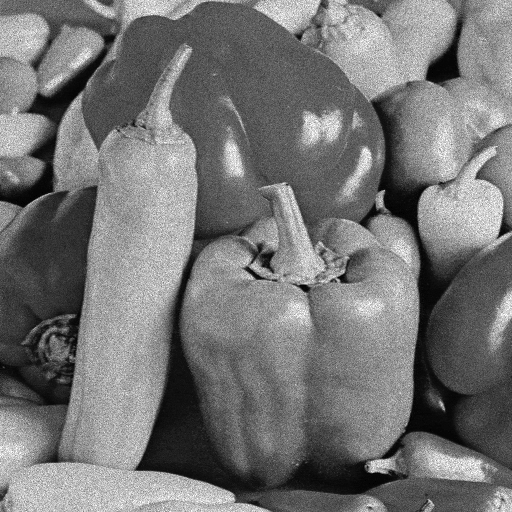}
\caption{Noisy Original\\ PSNR:25.97dB SSIM:0.67}
\label{fig:Peppers_noisy_orig}
\end{subfigure}
\centering
\begin{subfigure}[t]{0.24\linewidth}
\includegraphics[width=\linewidth]{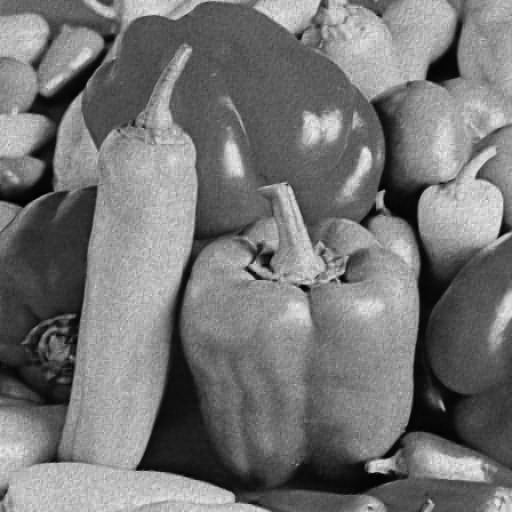}
\caption{SMoE-GD\\PSNR:30.56dB SSIM:0.76}
\label{fig:Peppers_noisy_GD}
\end{subfigure}%
\centering
\begin{subfigure}[t]{0.24\linewidth}
\includegraphics[width=\linewidth]{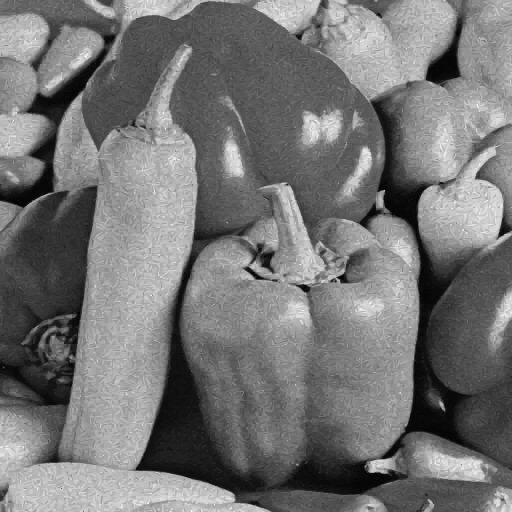}
\caption{SMoE-AE\\PSNR:29.86dB SSIM:0.73}
\label{fig:Peppers_noisy_AE}
\end{subfigure}%
\centering
\begin{subfigure}[t]{0.24\linewidth}
\includegraphics[width=\linewidth]{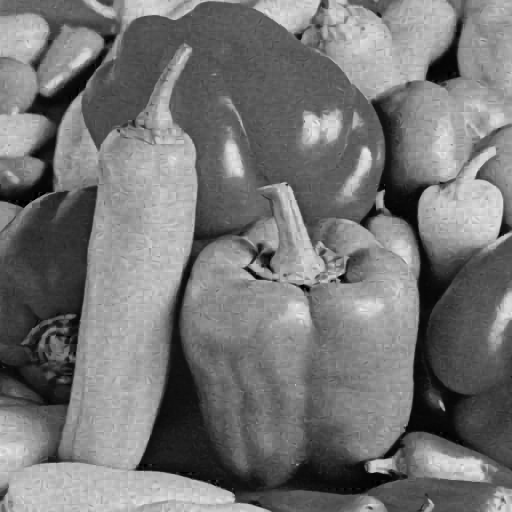}
\caption{SMoE-AE$_{noisy}$\\PSNR:30.94dB SSIM:0.77}
\label{fig:Peppers_noisy_AE}
\end{subfigure}%
\caption{Test image \textit{Peppers} with added speckle noise of variance $\delta^2=0.01$}
\label{fig:Peppers_noisy}

\end{figure}
The robustness to noise is examined and the autoencoder is trained on noisy images. Each $8\times8$ block is overlaid with additive speckle noise of variance $\delta^2 = 0.01$ before serving as input data to the encoder network and trained to calculate the loss over the original block. By adding noise to each block individually in every training iteration, the noise pattern changes moderately in each training epoch and successfully models real noise occurrences. The encoder network can adapt to the noise pattern to model the underlying edges in images, and subsequently denoise the images. The framework needs to be trained for every noise pattern separately.\\
%First, we showcase the robustness against noise of SMoE-AE without S-SMoE and present SMoEs native denoising capabilities. 
Tab. \ref{tab:added_noise} compares SMoE-GD and SMoE-AE for added speckle noise with a variance of $\delta^2=0.01$, as well as the effect of training SMoE-AE$_{noisy}$ on noisy images. SMoE is capable of performing light denoising tasks natively by creating an edge-aware model. Training the autoencoder on noise improves reconstruction quality for images with long and distinct high-contrast edges, like for the test image \textit{Cameraman} or \textit{Boats}. Images with a lot of high-frequency data like \textit{Baboon} or \textit{Bridge} do not benefit from SMoE-AE$_{noisy}$ to the same extent, since the encoder produces overly smooth blocks and therefore erases primary high-frequency data. Overall on the \textit{Kodak} data set the noise trained SMoE-AE$_{noisy}$ outperforms SMoE-GD by almost $1$dB. In Fig. \ref{fig:Peppers_noisy} the effect of additional smoothing is visible in the smooth pepper skin, which is notably more even in the reconstruction of SMoE-AE$_{noisy}$ compared to SMoE-GD and SMoE-AE. \\
Denoising results using the S-SMoE approach are provided in Tab. \ref{tab:trainedOnNoise} employing SMoE-AE$_{noisy}$, omitting the SMoE-GD comparison due to excessive run-time demands. Combining the edge-aware SMoE model with averaging over multiple hypotheses for each pixel in an image, results in great denoising improvements. For $S=1$ an improved PSNR of $1.98$dB and $0.07$ SSIM can be reported on the \textit{Kodak} data set in comparison to $S=8$. Test image \textit{Kodim23} from the \textit{Kodak} data set is depicted in Fig. \ref{fig:Kodak_noisy} for $S=1$ and $S=8$ and compared to the BM3D result. (best viewed electronically and zoomed in) For $S=1$ blocking artifacts are invisible and the noise is drastically reduced. The denoising results are comparable to state-of-the-art BM3D \cite{dabov2007image} approach.
\begin{figure}
\begin{subfigure}[t]{0.24\linewidth}
\includegraphics[width=\linewidth]{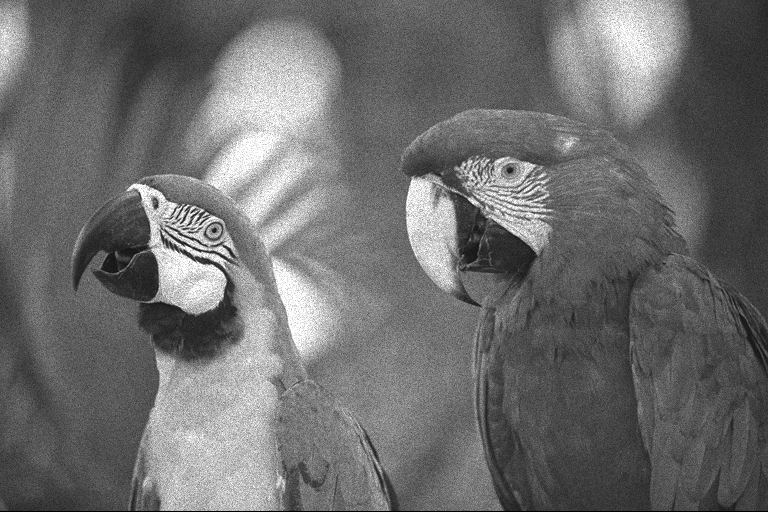}
\caption{Noisy Original\\ PSNR:26.78dB SSIM:0.53}
\label{fig:Kodak_noisy}
\end{subfigure}%
\centering
\begin{subfigure}[t]{0.24\linewidth}
\includegraphics[width=\linewidth]{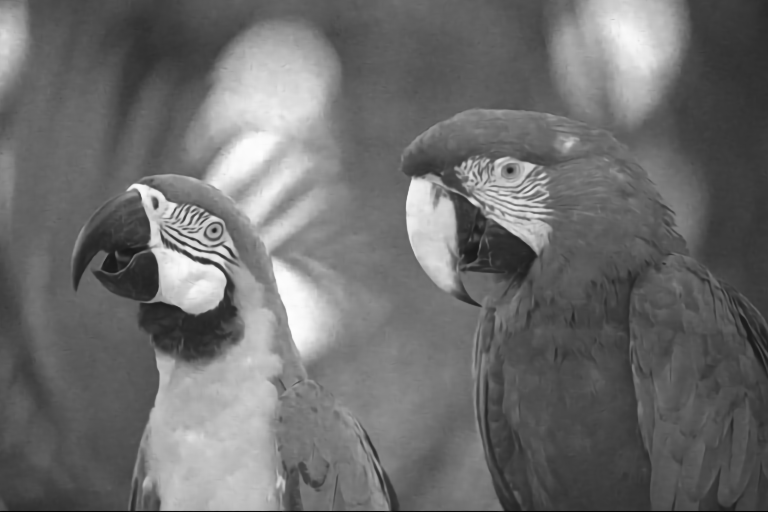}
\caption{S-SMoE $S=1$\\ PSNR:35.12dB SSIM:0.93}
\label{fig:Kodak_noisy_1}
\end{subfigure}%
\centering
\begin{subfigure}[t]{0.24\linewidth}
\includegraphics[width=\linewidth]{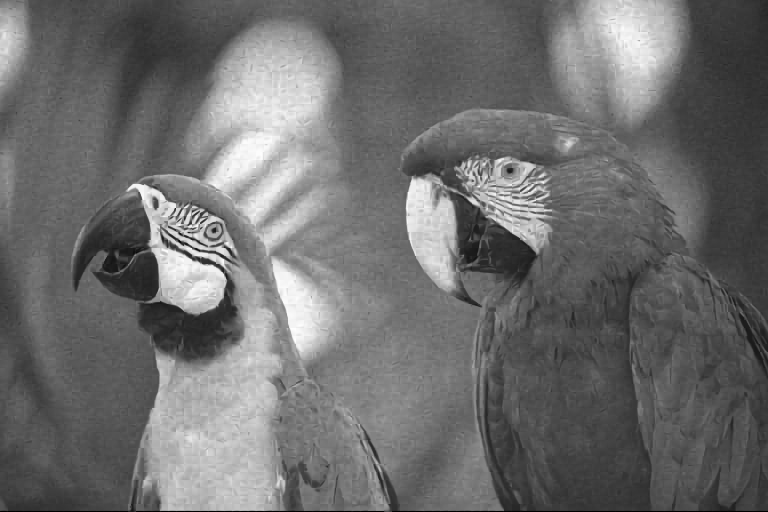}
\caption{S-SMoE $S=8$\\ PSNR:32.43dB SSIM:0.84}
\label{fig:kodak_orig}
\end{subfigure}
\centering
\begin{subfigure}[t]{0.24\linewidth}
\includegraphics[width=\linewidth]{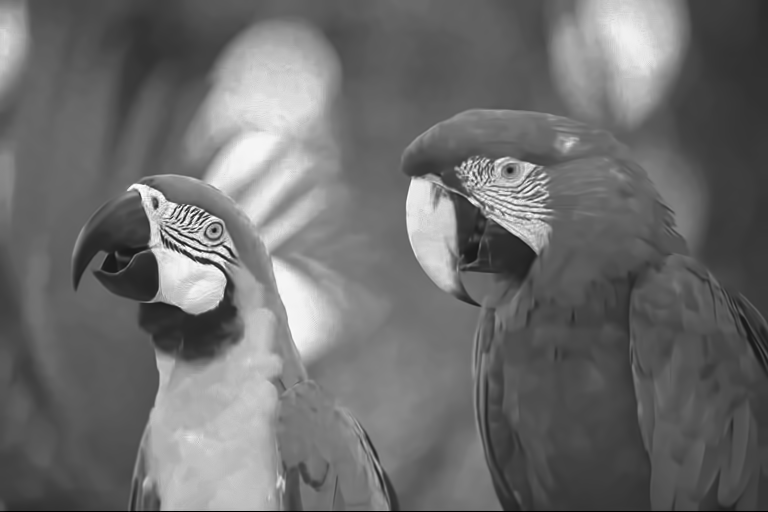}
\caption{BM3D Reconstruction\\ PSNR:35.1dB SSIM:0.92}
\label{fig:Kodak_noisy_8}
\end{subfigure}%
\caption{Test image \textit{Kodim23} with added speckle noise of variance $\delta^2=0.01$}
\label{fig:Kodak_noisy}
\end{figure}
\begin{table}[]
\centering
\caption{S-SMoE denoising results with SMoE-AE$_{noisy}$}
\label{tab:trainedOnNoise}
\begin{tabular}[t]{l l l l l l l l l}

\hline
&&& \multicolumn{6}{c}{S-SMoE with SMoE-AE$_{noisy}$}\\
&\multicolumn{2}{c}{Speckle noise $\delta^2=0.1$}&\multicolumn{2}{c}{$S=1$}&\multicolumn{2}{c}{$S=4$}&\multicolumn{2}{c}{$S=8$}\\
\hline
Sequence&PSNR [dB]&SSIM&PSNR [dB]&SSIM &PSNR [dB]&SSIM&PSNR [dB]&SSIM\\
%&[dB&&[bpp]&dB&&[bpp]&dB&\\
Baboon&25.53&0.73&29.27&0.87&28.54&0.86&27.10&0.82\\
Boats&25.38&0.62&31.38&0.85&30.83&0.84&29.60&0.81\\
Bridge&26.35&0.81&27.47&0.83&27.04&0.82&26.07&0.78\\
Cameraman&25.64&0.56&35.15&0.93&34.33&0.92&32.60&0.90\\
Elaine&25.09&0.56&32.04&0.78&31.70&0.77&30.97&0.75\\
Lena&25.69&0.56&34.15&0.90&33.52&0.89&32.16&0.86\\
Livingroom&25.95&0.67&31.36&0.87&30.77&0.86&29.51&0.83\\
Peppers&25.97&0.61&32.89&0.92&32.46&0.83&31.49&0.81\\
Kodak&26.90&0.67&31.01&0.87&30.50&0.86&29.03&0.80\\
\hline
Encoding Time [s]&\multicolumn{2}{c}{n/a}&\multicolumn{2}{c}{41.60}&\multicolumn{2}{c}{3.00}&\multicolumn{2}{c}{0.70}\\
Decoding Time [s]&\multicolumn{2}{c}{n/a}&\multicolumn{2}{c}{13.95}&\multicolumn{2}{c}{0.88}&\multicolumn{2}{c}{0.24}\\
%&0.16&25.15&0.64&-&-&-&0.16&\textbf{25.54}&\textbf{0.68}\\
\hline
\end{tabular}

\end{table}

\section{Conclusion}
In this paper, we proposed an advanced autoencoder framework to predict Steered Mixture-of-Experts parameters for improving reconstruction quality for complex and noisy images, while reducing run time demands drastically compared to gradient descent optimized SMoE models. Prior autoencoder implementations were restricted to radial kernels, consequently providing narrow applicability with other SMoE applications and limiting the complexity of the reconstructed images. This work utilizes steering capabilities of kernels and reduced block size and achieved higher reconstruction qualities in both objective and subjective measures, simultaneously maintaining run-time savings in contrast to iterative gradient descent optimization. Furthermore, the robustness to noise by training the encoder network on noisy ground truth images, makes it applicable to denoising applications using SMoE models.\\
Further research is necessary to explore adaptive block sizes and kernel numbers, to fully benefit of SMoEs ability to model long-range correlation in images.

\bibliographystyle{ieeetr}          
\bibliography{bibliography}

\end{document}